\documentclass[showpacs,pra,preprint,groupedaddress]{revtex4-1}
\usepackage{amsmath} 
\usepackage{amssymb}
\usepackage{graphicx}
\usepackage{color} 
\usepackage{url}
\usepackage{hyperref}
\begin{document}
\title{Quantum phase transition of two-mode Bose-Einstein
condensates with an entanglement order parameter}
\author{Wei~Fan} 
\author{Yan~Xu}
\email{x1y5hot@gmail.com}
\altaffiliation{Centre for Quantum Technologies, National University of 
Singapore, 3 Science drive 2, Singapore 117543}
\author{Bing~Chen}
\affiliation{College of Science, Shandong University of Science and
Technology, Qingdao 266510, China}
\author{Zhaoyang~Chen}
\email{zychen@berkeley.edu}
\affiliation{Department of Mechanical Engineering, University of California,
Berkeley, CA 94720, USA}
\author{Xunli Feng}
\author{C H Oh}
\email{phyohch@nus.edu.sg}
\affiliation{Center for Quantum Technologies and Physics Department,
    Faculty of Science,
        National University of Singapore,
            2 Science Drive 3,
                Singapore 117542 
                    Republic of Singapore
                    }
\date{\today}

\begin{abstract}
        The ground state entanglement of the two-mode Bose-Einstein condensate is 
	investigated through a quantum phase transition approach. The 
	entanglement measure is taken as the order parameter and this is a
        non-local order parameter, which is different from the conventional order
        parameter of the Mott insulator-superfluid phase transitions. For
        this non-local order parameter, scaling behavior corresponding to a continuous 
	phase transition  is obtained  and a power-law divergence 
	near the critical region follows it. This scaling behavior of quantum
        entanglement is analyzed by the finite-size scaling and the critical
        exponents are  obtained as $\nu = 1.01$ and $\gamma = 0.86$. A close
        connection between quantum fluctuations and the phase transition of
        entanglement is also obtained. 
\end{abstract}
\pacs{03.75.Gg, 05.30.Rt}
\maketitle

\section{Introduction}

Quantum entanglement is a key feature of quantum information theory and
it is one of the most active research areas in recent
years\cite{RevModPhys.81.865}, especially in  the area of its combination with condensed
matter systems\cite{RevModPhys.80.517}. Beyond its
generation and application, one of the essential
questions  is how to understand the process that happens in a physical system when
it transits from
non-entangled states to entangled states. One approach to study this
phenomenon  is to treat it as a quantum phase transition, where the order
parameter is  the entanglement measure.
This approach combines the theory of critical phenomenon 
with the theory of quantum information. For spin lattice models, many
results have been obtained.  The 
concurrence~\cite{PhysRevA.54.3824,PhysRevLett.80.2245} is often used as the 
entanglement measure in spin models. Entanglement exhibits scaling behavior
near the critical region and the critical behavior is shown to be 
logarithm~\cite{nature416.608, PhysRevLett.97.220402, PhysRevLett.93.086402,  
PhysRevLett.95.196406}. The phase transition is second-order for the ferromagnetic 
case and is first-order for the antiferromagnetic case~\cite{PhysRevA.69.054101}.  

While the critical behavior of quantum entanglement in  spin 
models have widely been studied, there are very few studies on that in  boson 
systems.  It is thus of interest to investigate the critical behavior of 
quantum entanglement in boson systems. 

One extensively studied  boson system  in quantum entanglement is the two-mode
Bose-Einstein condensates coupled via Josephson tunneling
\cite{PhysRevA.67.013609}. It is described by the Hamiltonian~\cite{RevModPhys.73.307}
\begin{equation}
\label{hamiltonian}
\mathcal{H} = \frac{K}{8} (N_1 - N_2)^2 - \frac{\Delta \mu}{2} (N_1 - N_2) -
\frac{\varepsilon_J}{2}(a_1^\dagger a_2 + a_2^\dagger a_1),
\end{equation}
where $a_1, a_2$ are the annihilation operators for the two modes (1 and 2)
respectively, and $N_1 = a_1^\dagger a_1, N_2 = a_2^\dagger a_2$ are the
corresponding number operators. The parameter $K$ provides the atom-atom
interaction, $\Delta \mu$ is the difference in the chemical potential between
the two modes and $\varepsilon_J$ is the coupling for tunneling. This Hamiltonian
describes both the double-well Bose-Einstein condensate and the two-level
Bose-Einstein condensate in a single potential. For the first case, the
tunneling between the two wells must be small to use this Hamiltonian, while
for the second case, there is no such restriction. We
will show in this paper that the phase transition occurs at very small couplings, so
the quantum phase transition approach can describe both cases. 
The entanglement production in this system has been extensively studied
~\cite{PhysRevA.67.013609,PhysRevA.71.013601,Nature.409.63,e2008-00472-6, ISI:000227261000008}.
The von Neumann entropy \cite{neumann1995} $E(\rho)$ is the usually used entanglement 
measure, where $\rho$ is the density matrix of the system, and for a system
size of $N$ particles, the maximum entropy is $E_{max}=\log_2(N+1)$. 

This Hamiltonian~\eqref{hamiltonian} is in fact a two-site version of the
Bose-Hubbard model~\cite{PhysRevB.40.546}. When varying the ratio between the
interaction term the coupling term through
a critical value, a quantum phase transition occurs in the Bose-Hubbard
model, which is the Mott insulator to the superfluid
transition~\cite{RevModPhys.80.885}. 
This phase transition is driven by quantum fluctuations and
the order parameter is the conventional wave function.  In the Mott insulator
phase, atoms are localized in  lattice sites, while in the superfluid phase,
atoms spread out over the whole system. 
Although  the insulator-superfluid phase transition is studied
extensively~\cite{PhysRevB.58.R14741,Greiner_Mandel_Esslinger_Hänsch_Bloch_2002,Elstner_Monien_1999,     
PhysRevLett.80.2189,PhysRevLett.97.200601} both in theory and in experiment,
it is interesting to investigate what would happen to the Bose-Hubbard model
when taking a non-local order parameter, rather than the conventional order
parameter. 

 In this paper, we present such a study for the simplest two-site
Bose-Hubbard system, the two-mode Bose-Einstein condensate. The entanglement
measure, the von Neumann entropy,  is taken as the non-local order parameter. 
We show that there is a critical point and entanglement exhibits scaling
behavior near the critical point, which can be analyzed using the theory of
critical phenomena. We 
identify this as a continuous phase transition. This phase transition is
different from the insulator-superfluid phase transition, because it is
obtained for a non-local order parameter, rather than the conventional order
parameter. The critical behavior of quantum entanglement is shown to be
power-law divergent, which is different from the logarithm divergent of spin
lattice models.  Our work may help the combination of methods in critical phenomena
and quantum information theory for the boson systems, especially for the
Bose-Hubbard model. Further extension of this work is to investigate 
quantum phase transitions in the Bose-Hubbard model of dozens of lattice
site, where a new entanglement measure is also needed to be put up.

\section{Continuous Phase Transition}

In this paper, we only consider the case $K > 0$, corresponding to a
repulsive interaction between atoms. The total particle number is conserved
and we set $\Delta \mu =0$. Using the angular momentum operators 
\begin{eqnarray*}
J_z &=& \frac{1}{2} (N_2 - N_1) \\
J_x &=& \frac{1}{2} (a_1^\dagger a_2 + a_2^\dagger a_1) \\
J_y &=& \frac{i}{2} (a_1^\dagger a_2 - a_2^\dagger a_1)
\end{eqnarray*}
and neglecting constant terms, the Hamiltonian~\eqref{hamiltonian} is
rewritten as 
\begin{equation}
  \label{eq:1}
  \mathcal{H} = \chi J_z^2 - \Omega J_x ,
\end{equation}
where $\chi = K/2$ and $\Omega = \varepsilon_J$. As we are only interested in
the ratio between the two competing energy terms, it is convenient to
introduce the dimensionless parameter $\Omega/\chi$ in the
calculation, so the Hamiltonian can be reduced to 
\begin{equation}
  \label{eq:2}
  \mathcal{H} = J_z^2 - \Omega J_x, 
\end{equation}
where we have redefined $\Omega$ using the dimensionless parameter, i.e.,
$\Omega/\chi \rightarrow \Omega$. This dimensionless coupling parameter can be viewed as
an 'external field' by analogy with Ising models. We also define the dimensionless entropy
$E(\rho)/E_{max} \rightarrow E(\rho)$ to make it easier to compare the
results of different system sizes.
We use numerical diagonalization  to calculate \cite{matplotlib,scipy} the
ground state entanglement and its susceptibility with respect to the external
field $\Omega$. 

\begin{figure} 
\begin{center} 
\includegraphics[width=16cm]{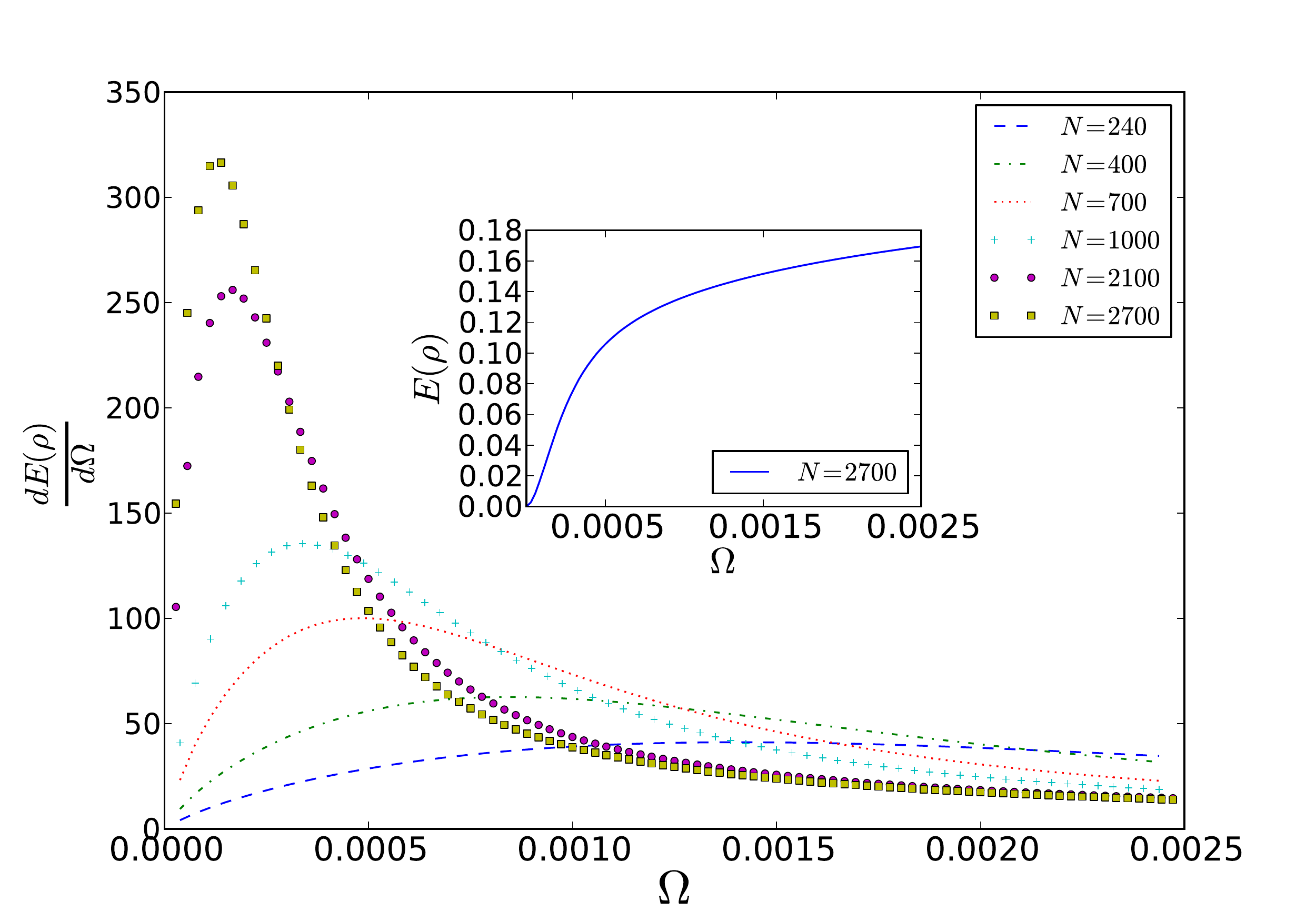}
\caption{(color online) The susceptibility $\frac{d E(\rho)}{d \Omega}$ of the quantum entanglement
with respect to the external field $\Omega$ for system sizes of 
$N=240,400,700,1000,2100,2700$.
 A continuous phase transition occurs as the
susceptibility diverges with the system size. The critical point $\Omega_m$  
where the susceptibility  
attains its maximum $\frac{d E(\rho)}{d \Omega}_m$ lies in the Fock regime and 
this critical susceptibility
diverges with the system 
size. 
The inset depicts the change of the order parameter-the ground 
state entanglement for the system of $N=2700$, which increases continuously from 
zero. We choose even particle numbers because for odd particle numbers there 
is a degeneracy of the ground state when $\Omega=0$. 
} 
\label{figure0} 
\end{center}
\end{figure}

 We first calculate the susceptibility $\frac{d E(\rho)}{d \Omega}$ with 
 respect to the coupling $\Omega$ for various system sizes, which is shown in  
Fig. \ref{figure0}. We see that there is a critical point $\Omega_m$ for each 
system size, where the susceptibility reaches 
its critical value $\frac{d E(\rho)}{d \Omega}_m$.
The critical susceptibility $\frac{d E(\rho)}{d \Omega}_m$ increases with the 
system size and would be  
divergent for an infinite system size that corresponds to the thermodynamic 
limit, which implies 
that this is a continuous phase transition where there 
is no discontinuity in the order parameter, as depicted by the inset for the
system of $N=2700$ particles. This will be verified further in section \ref{finite-size}.

From Fig.~\ref{figure0}, the critical point $\Omega_m$ lies
in the small coupling regime, which means the phase transition 
occurs shortly after the external field is switched on. We could easily figure 
that the critical value is $\Omega_c = 0$ for an infinite system size of the 
thermodynamic limit. 
When $\Omega = 0$, the Neumann entropy is zero and there is no two-mode entanglement in the 
system; When $\Omega > 0$, the Neumann entropy gets a finite value and 
entanglement is generated in the system. That means the system transits from 
non-entangled states to entangled states, two essentially different states, 
once $\Omega$ is switched on from 0, so the critical value is just 0. This
will be verified further in section \ref{power-law}, where we numerically 
fit the critical point and the critical susceptibility for various system 
sizes. The critical point $\Omega_m$ is well fitted to $N$  by choosing 
$\Omega_c = 0$. 

\section{Power-law Divergence \label{power-law}}

\begin{figure} 
\begin{center} 
\includegraphics[width=16cm]{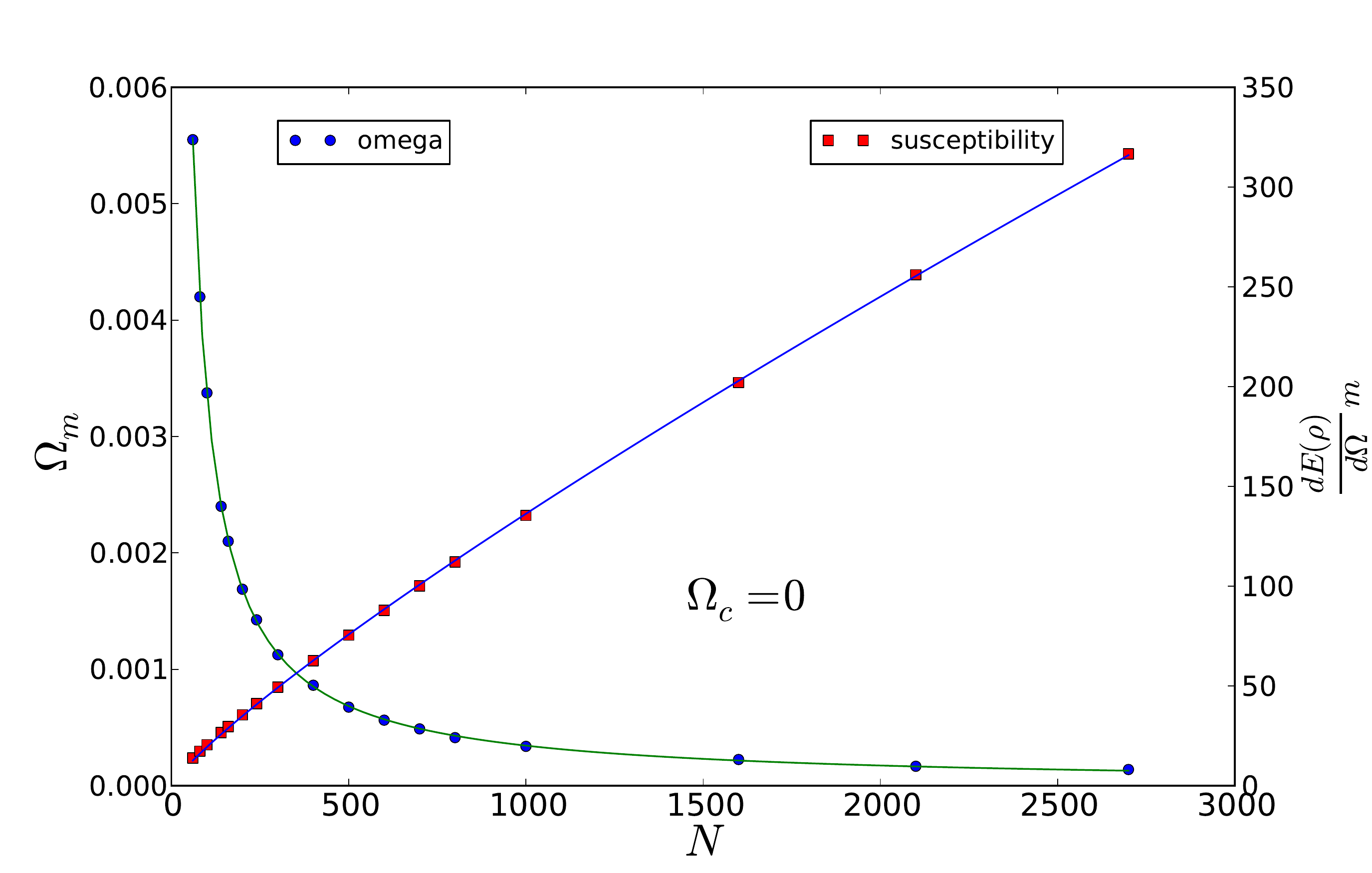}
\caption{(color online) The scaling behavior of the quantum phase transition with
the critical value $\Omega_c=0$.
The critical point $\Omega_m$ approaches 0 asymptotically by 
$\Omega_m \sim N^{-0.989062}$ and the susceptibility diverges in a power-law 
behavior captured by
$\frac{d E(\rho)}{d \Omega}_m \sim N^{0.846662}$, which is different from the 
logarithm divergence of spin lattice models.     
} 
\label{figure3} 
\end{center}
\end{figure}

The well-behaved relationship between the critical point and 
the system size in Fig.~\ref{figure3} is not just a coincidence. Actually it manifests the 
scaling behavior of quantum entanglement for this quantum system, which is 
typical in critical phenomenon.
From Fig.~\ref{figure3}, we obtain the scaling relationship 
\begin{equation}
	\label{critical-correlation}
\Omega_m =  0.319225N^{-0.989062}
\end{equation}
for the critical point and the scaling relationship 
\begin{equation}
	\label{critical-susceptibility}
	\frac{dE(\rho)}{d \Omega}_m=  0.393037 N^{0.846662}.
\end{equation}
for the critical susceptibility.
The scaling behavior of the susceptibility is power-law divergent, in contrast to the logarithm 
divergence of spin lattice systems~\cite{nature416.608}.  

This power-law divergence of the susceptibility can be understood in the thermodynamic limit
using a simple analysis. The basic idea is to truncate the Fock space of the
system to just three basis states and use them to approximate the state of
the system. The validity of this approximation lies in the fact that the
critical point is $\Omega_c=0$ and the delocalization process is very weak
near this critical point, which means that the transitions between different basis
states of the original Fock space are very weak  and we can use the three most important basis states for
approximation. This is verified at the end of the calculation in Eq.~\eqref{analytic-power},
where  a power-law behavior of the susceptibility is obtained  and the divergent exponent does
not differ much from that of the numerical simulation.     

We choose the Fock space basis $|N_1,N_2>$ for the system, 
where $N_1$ is the particle number in the first site and $N_2$ is the particle 
number on the second site. 
When $\Omega = 0$, the ground state is $|N/2,N/2>$ with energy $E=0$, that is, the system is in 
a self-trapping state without particle tunneling between the two sites. As 
$\Omega$ increases, the particles begin tunneling between the two sites and 
this delocalization process connects different basis states, so the system is 
described by $\sum_{n=0}^N c_n|n,N-n>$. 
The critical value is $\Omega_c = 0$ and the delocalization process is very 
weak near this region, so we can truncate the Fock space of the system to just 
three basis states $|N/2,N/2>$, $ |N/2-1,N/2+1>$ and $|N/2+1,N/2-1>$, then the state of 
the system is $|\psi> = c_0 |N/2,N/2> + c_1 |N/2-1,N/2+1> + c_2 |N/2+1,N/2-1>$, where 
we assume the coefficient $c_i$ to be real numbers for simplicity.
As the probabilities of tunneling between the two sites are equal, the 
coefficients $c_1$ and $c_2$ are equal. Combining with the normalization condition 
$c_0^2 + c_1^2 + c_2^2 =1$, we get the relationship $c_0 = \sqrt{1-2c_1^2}$. 
We next calculate the approximate ground state to determine the value of the coefficients,
\begin{eqnarray*}
	\mathcal{H}|\psi>&=&-\frac{\Omega N}{2}c_1 \{|\frac{N}{2},\frac{N}{2}> 
	+\frac{-\frac{\Omega N}{4}\sqrt{1-2c_1^2} + c_1}{-\frac{\Omega 
	N}{2}c_1}[|\frac{N}{2}-1,\frac{N}{2}+1> + |\frac{N}{2}+1,\frac{N}{2}-1>]\}\\
	&=&E|\psi>,
\end{eqnarray*}
where the approximation $\sqrt{N/2(N/2+1)}\sim N/2$ is taken. 
The critical point $\Omega_c=0$ determines that 
 $c_1$ is a small 
number. From $|\psi>=c_0[|N/2,N/2> + c_1/c_0(|N/2-1,N/2+1> + |N/2+1,N/2-1>)],$ 
we obtain
\[
E = -\frac{\Omega N}{2} \frac{c_1}{c_0},
\]
which is approximately zero and is the ground state energy near $\Omega_c=0$,
and 
\[
\frac{-\frac{\Omega N}{4}\sqrt{1-2c_1^2} + c_1}{-\frac{\Omega 
N}{2}c_1} = \frac{c_1}{\sqrt{1-2c_1^2}},
\]
which gives the value 
\[
c_1^2 = \frac{1}{4}( 1 - \frac{1}{\sqrt{1 + \frac{\Omega^2N^2}{2}}}).
\]
There are two values of $c_1^2$ and what we choose is the smaller one.
Substituting the values of the coefficients into the von Neumann entropy
\[
E(\rho) = -c_0^2 \log_2 c_0^2 - c_1^2 \log_2 c_1^2 - c_2^2 \log_2 c_2^2
\]
and taking its derivative with respect to $\Omega$ gives 
\begin{equation}
 \label{analytic-power}
	\frac{dE(\rho)}{d \Omega} \sim \frac{\Omega N^2}{\Omega^4N^4} \sim N^{0.97},
\end{equation}
where the relationship $\Omega N \sim N^{0.01}$ from
Eq.~\eqref{critical-correlation} in the thermodynamics limit is used.
Thus we briefly illustrate the power-law divergence of the susceptiblity in 
the thermodynamic limit. 

The divergent exponent obtained in the analytic calculation is 0.97 and it is
different from the value 0.85 of the numerical simulation in Eq.~\eqref{critical-susceptibility}. 
This difference may be accounted by the finite-size effects and the truncation
errors. First, the analytic calculation manifests the thermodynamic limit, where
there is no finite-size effect. The numerical result, however, is 
influenced by the finite-size effects, so this may be one of the reasons for the
difference between the divergent exponents. Second, we adopt approximation in
the analytic calculation by truncating the Fock space of the system to just
three basis states. The numerical simulation, however, includes the full Fock
space. The neglected basis states would certainly contribute to the result,
although their amplitudes are small near the critical point. So the difference between the
divergent exponent is also influenced by the truncation errors.

\section{Finite-size Scaling\label{finite-size}}

\begin{figure} 
\begin{center} 
\includegraphics[width=16cm]{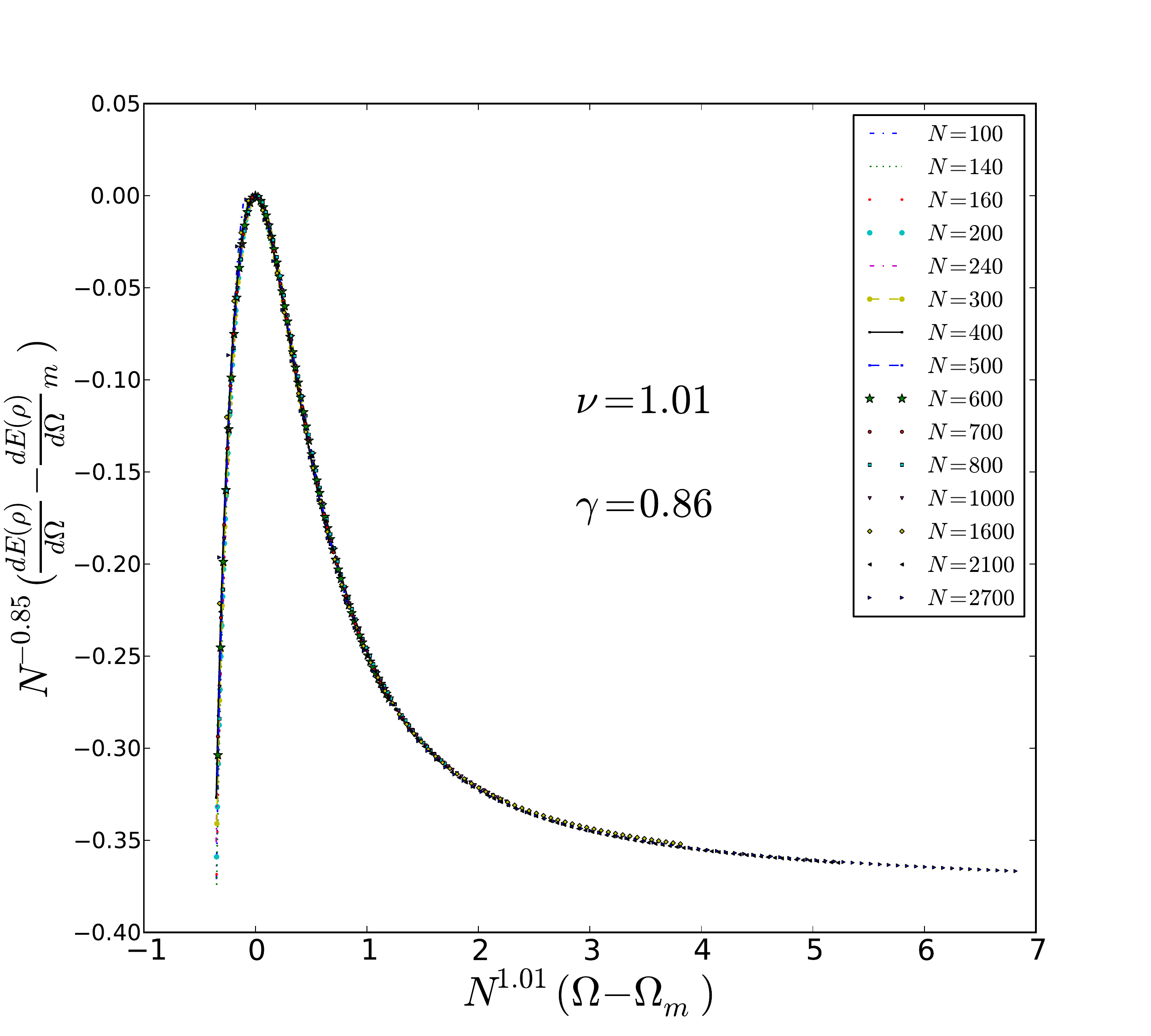}
\caption{(color online) The finite-size scaling for the quantum phase transition of quantum entanglement. 
After the susceptibility is reduced by the system size to 
$N^{-0.85}(\frac{d E(\rho)}{d \Omega} - \frac{d E(\rho)}{d \Omega}_m)$,
it becomes a function of $N^{1.01}(\Omega-\Omega_m)$. 
Data from a broad range of system sizes are collected on this single curve. 
The critical exponent obtained is $\nu=1.01$ and $\gamma=0.86$.
} 
\label{figure4} 
\end{center}
\end{figure}

A key feature of the critical phenomenon is the finite-size scaling.
Phase transitions only occur at the thermodynamic limit, while numerical 
simulations can only deal with finite system sizes. To extract information 
from results obtained from the finite system, the finite-size scaling is 
required, where the 
effect of finite system sizes are eliminated by collecting all 
data of various system sizes onto a single curve and the critical exponent can 
be deduced in this process. 
In the phase transition of thermal order parameters, e.~g.~, the 
magnetization, the critical exponent $\nu$ of the correlation length satisfies 
$|T - T_c| \sim N^{-1/\nu}$. By analogy, we obtain
$\nu = 1/0.989062 \sim 1.01$ from Eq.~\ref{critical-correlation}, which is the critical 
exponent for the quantum phase transition of quantum entanglement.
This critical exponent gives the reduced coordinate 
$N^{\nu}(\Omega - \Omega_m)$ for all the finite system sizes. From 
Eq.~\ref{critical-susceptibility}, the susceptibility is reduced to 
$N^{-0.85}(\frac{d E(\rho)}{d \Omega} - \frac{d E(\rho)}{d \Omega}_m)$. If the 
quantum entanglement of this model manifests quantum phase transitions, then 
all data of various system sizes could be collected onto a single cure using 
the above reduced coordinates. This is indeed the case as
exhibited in Fig. \ref{figure4}. Again this relationship is not just a 
coincidence. It illustrates that quantum entanglement of this model indeed belongs 
to critical phenomenon. Resorting to the the phase transition of the 
Magnetization, where the susceptibility $\chi$ of the Magetization is reduced 
to $N^{-\gamma/\nu}\chi$, we obtain the critical exponent 
$\gamma = 0.85\nu \sim 0.86$ in this model.

\section{Quantum Fluctuations}

\begin{figure} 
\begin{center} 
\includegraphics[width=16cm]{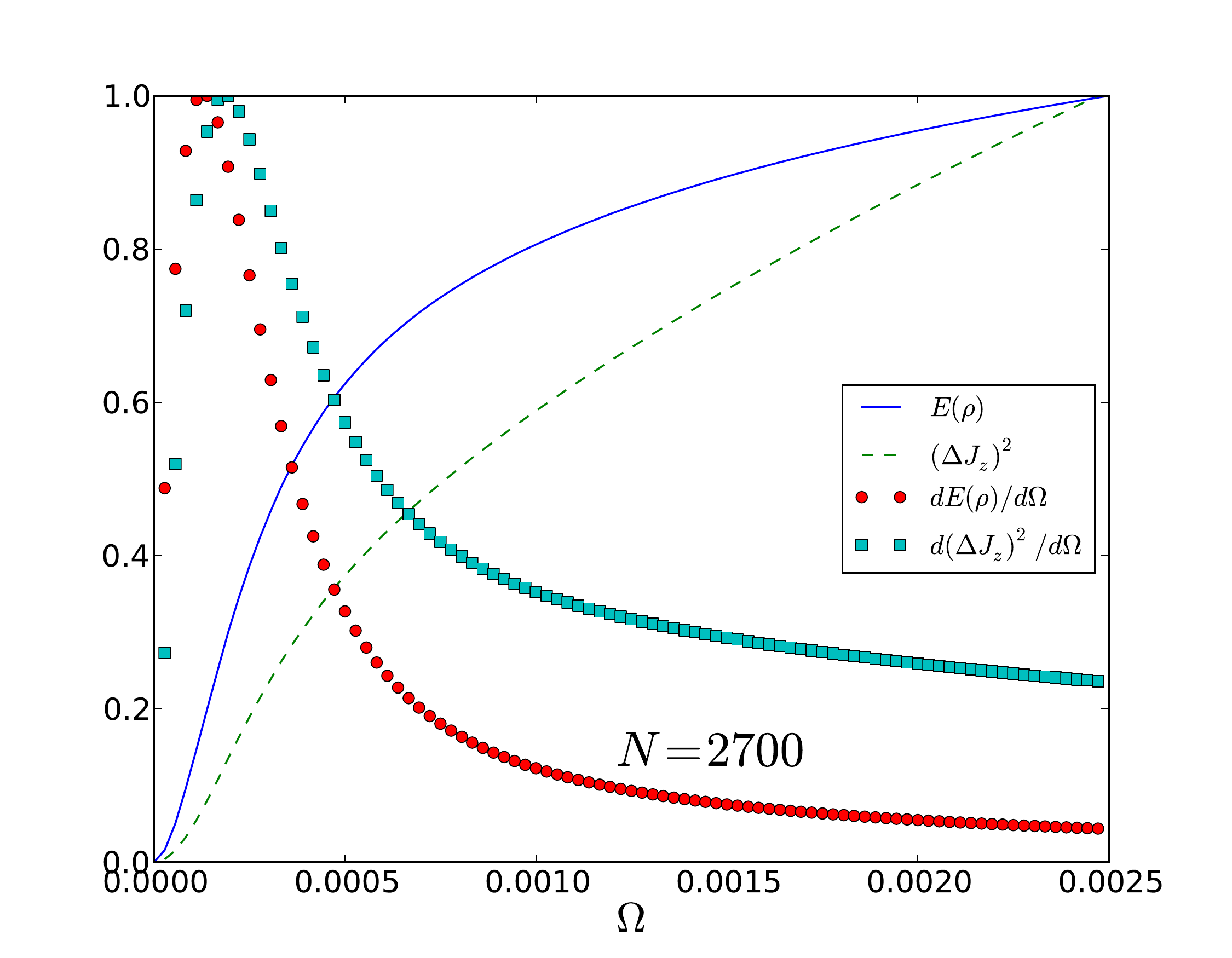}
\caption{(color online) Comparison between $(\Delta J_z)^2$, $E(\rho)$ and their derivatives 
with respect to $\Omega$. 
Their values are reduced to $1$ by their maximum values.  The increase of the 
quantum fluctuation corresponds to the increase of the order parameter, which
indicates its connection with the phase transition of entanglement.  
There is a small 'delay' between the derivative of the fluctuation and the 
susceptibility, with the susceptibility obtaining maximum value first. This 
'delay' comes from the finite-size effects.
} 
\label{figure2} 
\end{center}
\end{figure}

The Mott insulator-superfluid transition is driven by
quantum fluctuations, which is common for quantum phase transitions. Here
we show that  a close connection also exists between quantum fluctuations and
the phase transition of entanglement .    In the dynamical regime of entanglement
production, the system  is required to undergo a delocalization process, where large quantum 
fluctuation exists, to generate entanglement.  So the quantum phase 
transition of entanglement should be closely related to quantum fluctuations.  In 
the angular momentum  
representation $|j,j_z>$, where $j=N/2$ and $j_z=-N/2,-N/2+1,\ldots,N/2$, the 
quantum fluctuation is $(\Delta J_z)^2=\langle J_z^2 \rangle - \langle J_z \rangle^2$.
We show that $(\Delta J_z)^2$ and $E(\rho)$  have a similiar behavior, which
indicates their close connection with each other. 
 We plot $(\Delta J_z)^2$, $E(\rho)$ and their 
derivatives with respect to $\Omega$ in their reduced value in Fig.~\ref{figure2}. 
We see that both the quantum fluctuation and the quantum entanglement grows 
with the external field $\Omega$, and their growth corresponds to each other, 
which can be seen from their derivatives.
As quantum entanglement is a 
non-classical correlation, it is consistent that its quantum phase
transition is closely related to quantum fluctuations. 

There is a 'delay' between the derivative of $E(\rho)$ and that of 
$(\Delta J_z)^2$, where the derivative of $E(\rho)$ reaches its maximum value earlier 
than the derivative of $(\Delta J_z)^2$.   This is due to the finite-size 
effects. We are not working in the thermodynamic limit, so the derivatives 
between the quantum fluctuation and the quantum entanglement are not in 
complete correspondance. This is further confirmed by Table \ref{table1}, 
where the 'delay' $\Delta \Omega$ between the maximum points of the derivatives 
are calculated 
for various system sizes. We see that the 'delay' between them is decreasing 
as the system size growes, so we can figure that in the thermodynamic limit, 
the growth behavior of the entanglement and that of the fluctuation will approximately 
correspond to each other. 

\begin{table}
\caption{\label{table1} The 'delay' $\Delta \Omega$ between the 
maximum points of the derivatives of $E(\rho)$ and $(\Delta J_z)^2$ 
for various system sizes. As the system size increases, the 'delay' 
decreases, which means the growth behavior of the entanglement 
and that of the fluctuation are more closely related. This suggests 
that in the thermodynamic limit, the two growth behaviors will 
correspond to each other.}
\begin{ruledtabular}
	\begin{tabular}{c|c|c|c|c|c|c}
		N & 200 & 400 & 600 & 800 & 1600 & 2700\\
		\hline
		$\Delta \Omega$ & 0.000675 & 0.000375 & 0.000263 & 0.000188 & 0.000075 & 0.000055\\
	\end{tabular}
\end{ruledtabular}
\end{table}

\section{Summary}

In summary, we have studied the entanglement of a boson system from the 
quantum phase transition approach.  It is shown that in this system there is
 a continuous phase transition for the non-local order parameter   and 
entanglement exhibits scaling behavior near the critical point, with the 
critical exponent calculated to be $\nu = 1.01$ and $\gamma = 0.86$.  The critical 
behavior under discussion is different from that of the spin lattice models, 
because a power-law 
divergence is obtained for the boson system while it is logarithm divergence for the spin 
models. 
A further study of this phenomenon may consist of putting up an entanglement 
measure for Boson systems of more lattice sites, investigating the quantum 
phase transition of the Bose-Hubbard model of more lattice sites 
and obtaining its universality class. The renormalization group method specially
 for taking into account the effect of quantum
  entanglement~\cite{PhysRevLett.69.2863,1019601218492,PhysRevA.81.062313} may be used in that case.

\begin{acknowledgments}
Project is partly supported by the National Natural Science Foundation of
China (Grant No.~11105086), the National Research Foundation and Ministry of
Education, Singapore(Grant No.~WBS: R-710-000-008-271), the Natural Science
Foundation of Shandong Province (Grant No.~ZR2009AM026, BS2011DX029), the
basic scientific research project of Qingdao(Grant 
No.~11-2-4-4-(6)-jch), the basic scientific research business expenses of the
central university, and the Open Project of Key Laboratory for Magnetism and
Magnetic Materials of the Ministry of Education,Lanzhou University. 
\end{acknowledgments}

\bibliography{phys-rev-e}

\end{document}